\documentclass[]{pasj02}
\usepackage[switch,mathlines]{lineno} % add line number to manuscript
\usepackage{xcolor}

\jyear{2024}
\Received{}%{yyyy/mm/dd}
\Accepted{}%{yyyy/mm/dd}
\begin{document}

\title{Detection of the solar internal flows with numerical simulation and machine learning}

%%% begin:list of authors
% Do NOT capitalize all letters in "textsc".
\author{H. \textsc{masaki},\altaffilmark{1}\altemailmark\orcid{0009-0001-0858-6208} \email{afpa6720@chiba-u.jp} 
and
H. \textsc{hotta},\altaffilmark{2}$^{,\dag}$\orcid{0000-0002-6312-7944} \email{hotta.h@isee.nagoya-u.ac.jp}
% C-Firstname \textsc{C-Familyname},\altaffilmark{3}\altemailmark \email{ccccc@xxx.xxx.xx.xx}
% and 
% D-Firstname \textsc{D-Familyname}\altaffilmark{2}\altemailmark\orcid{0000-0000-0000-0000} \email{ddddd@xxx.xxx.xx.xx}}
}
%\footnotetext[$\dag$]{Present address: ....}

%%% end: list of authors

%% !!! Select 3 to 5 words from PASJ's keywords!!!
%% List of Key Words: https://academic.oup.com/pasj/pages/Pasj_Keywords
%% "\KeyWords{ }" always has to be placed before ``\maketitle''

\maketitle

\begin{abstract}
The solar interior is filled with turbulent thermal convection, which plays a key role in the energy and momentum transport and the generation of the magnetic field. The turbulent flows in the solar interior cannot be optically detected due to its significant optical depth. Currently, helioseismology is the only way to detect the internal dynamics of the Sun. However, long-duration data with a high cadence is required and only the temporal average can be inferred. To address these issues effectively, in this study, we develop a novel method to infer the solar internal flows using a combination of radiation magnetohydrodynamic numerical simulations and machine/deep learning. With the application of our new method, we can evaluate the large-scale flow at 10 Mm depth from the solar surface with three snapshots separated by an hour. We also apply it to the observational data. Our method is highly consistent with the helioseismology, whereas the amount of input data is significantly reduced.
\end{abstract}

%\pagewiselinenumbers
 \clearpage %
\section{Introduction}
The solar interior is filled with turbulent thermal convection. Thermal convection has several roles in solar dynamics. It transports the energy and momentum to construct the structure and the large-scale flows in the solar interior, respectively \citep{Hotta+2023}. Moreover, thermal convection is the origin of the solar magnetic field, which is the source of solar activities, such as sunspots, flares, and coronal mass ejections. On the surface, the turbulent motion causes the Poynting flux, which heats the solar corona and drives the solar wind \citep{CranmervanBallegooijen2005}.
\par
There have been many solar observations to date to evaluate the thermal convection on the solar surface \citep{bellot_rubio_2019LRSP...16....1B}. The details of the thermal convection on the solar surface are well understood. There is granulation with a typical scale of 1 Mm and supergranulation with a typical scale of 30 Mm. However, optical observations are restricted to the surface, and we cannot access any information on the solar interior due to the significantly large optical depth.
\par
We can currently access the solar internal dynamics only by using the helioseismology method \citep{Duvall+1993}. In particular, the acoustic waves are excited by the turbulent convection on the surface, and the wave propagates through the solar interior and emerges at the surface again. The oscillation observed on the surface conveys information about the solar interior. Helioseismology has revealed several essential phenomena in the solar interior, such as the precursor of sunspot emergence \citep{IlonidisZhaoKosovichev2011} and the large-scale flows in the deep convection zone \citep{Zhao+2013}. Although helioseismology is a powerful tool, it still has significant limitations. This method requires long temporal data with a high cadence. Because of the nature of the stochastic excitation of the wave, the observational data include a large amount of noise. To reduce the noise and detect the flow at 3-4 Mm depth, we typically require 12 h-long data with one minute cadence, i.e., 720 snapshots \citep{Sekii+2007}. This large amount of data increases the computational cost of helioseismology. In addition, the long-time data only provide the time-averaged result. The time scale for the near-surface layer (10 Mm) is several hours, and helioseismology cannot detect the temporal evolution of the layer.
\par
Meanwhile, radiative magnetohydrodynamic (RMHD) simulation has been significantly improved in the last two decades to understand solar convection dynamics, and nowadays, it can reproduce the solar flows and magnetic fields quite well \citep{SteinNordlund1998, Vogler+2005, Rempel+2009, HottaIijimaKusano2019, HottaToriumi2020}.
%The reproduction level for the deep thermal convection is still under debate \citep{Hotta+2023}. The observational
%result has not been converged \citep{Hanasoge+2012, Greer+2015}, and one of the results is significantly different from the simulation one. On the other hand,
These simulations are qualitatively consistent with the observations \citep{NordlundSteinAsplund2009}, and the simulations for the small-scale magnetic field also match the observations \citep{Rempel2014}. There is an indication that we can reproduce the thermal convection properties in the layer shallower than 20 Mm \citep{Lord+2014}.
\par
Machine learning is an optimization method for mathematical models that is now applied in various fields. By recognizing patterns from large amounts of data, machine learning effectively solves complex problems that were difficult, improving prediction accuracy and computational efficiency, and achieving excellent outcomes across many fields.
Machine learning has also been applied to solar thermal convection. DeepVel estimates horizontal thermal convection on the solar surface from two images taken by 30 seconds \citep{Asensio+2017}. This method is superior to other techniques for estimating horizontal flows, such as Local Correlation Tracking \citep{Tremblay+2018}. Additionally, \cite{Tremblay+2020} improves DeepVel's performance with a U-net architecture. \cite{Ishikawa+2022} examines prediction performance at different spatial scales and suggests ways to enhance machine learning's performance in small spatial scales. In \cite{masaki_2023PASJ...75.1168M}, the amount of observational data used for estimation is reduced, enabling predictions from snapshots.
\par
In this study, we propose a novel method to infer the solar internal dynamics taking advantage of the well-established RMHD simulations and the fast-growing machine learning technique. The RMHD simulation in the near-surface is fairly reliable, and the neural network is trained based on our simulation result. Our method is also validated with the observation data taken with the SDO/HMI (Solar Dynamics Observatory/Helioseismic and Magnetic Imager) satellite \citep{PesnellThompsonChamberlin2012}.
\par
\section{Method}
\subsection{RMHD simulation}
We use the radiation magnetohydrodynamics simulation code R2D2\footnote{Radiation and RSST for Deep Dynamics; where RSST stands for the reduced speed of sound technique \citep{hotta_2012A&A...539A..30H} which we turned off in this study.} \citep{HottaIijimaKusano2019,HottaIijima2020} in this study, to calculate the internal convective velocity for training data. R2D2 solves the following MHD equations:

\begin{eqnarray}
  \frac{\partial\rho}{\partial t}
  &=&
  -\nabla\cdot(\rho \boldsymbol{v}),
\\
  \frac{\partial }{\partial t}(\rho \boldsymbol{v})
  &=&
  -\nabla\cdot(\rho \boldsymbol{v}\boldsymbol{v})
  -\nabla p
  +\rho \boldsymbol{g}
  +\frac{1}{4\pi}
  \left(\nabla\times\boldsymbol{B}\right)\times \boldsymbol{B},
\\
  \frac{\partial \boldsymbol{B}}{\partial t}
  &=&
  \nabla
  \times\left(\boldsymbol{v}\times\boldsymbol{B}\right),
\\
  \rho T \frac{\partial s}{\partial t}
  &=&
  \rho T \left(\boldsymbol{v}\cdot\nabla\right)s + Q,
\\
  p &=& p\left(\rho,s\right),
%\end{align}
%\end{equation}
\end{eqnarray}
where $\rho$, $\boldsymbol{v}$, $p$, $T$, $\boldsymbol{g}$, $\boldsymbol{B}$, $s$, and $Q$ represent the density, velocity, pressure, temperature, gravitational acceleration, magnetic field, entropy, and radiative heating, respectively. The pressure $p$ is calculated from a table prepared using the OPAL equation of state \citep{rogers_1996ApJ...456..902R}. The equation is solved with a fourth-order Runge-Kutta method for the time derivatives and a fourth-order accuracy for the spatial derivatives. The radiative heating $Q$ is solved using a grey approximation. In this study, we solve only up- and downward energy transfer to reduce computational costs. See \cite{HottaIijima2020} for more details.
\par
The computational domain is 49.152 Mm horizontally in each direction and 24.576 Mm vertically. The top boundary is located 700 km above the surface of the Sun, i.e., an optical depth of $\tau=1$. The horizontal grid spacing is set to 96 km. We use nonuniform grid spacings for the vertical direction. Around the photosphere, the grid spacing is 48 km and gradually increased to about 90 km around the bottom boundary. As a result, the simulation domain is solved with $384\times512\times512$ grid points. We continue each simulation for about two weeks. The data output cadence is five minutes.
\par
\begin{figure*}
  \begin{center}
   \includegraphics[width=\textwidth]{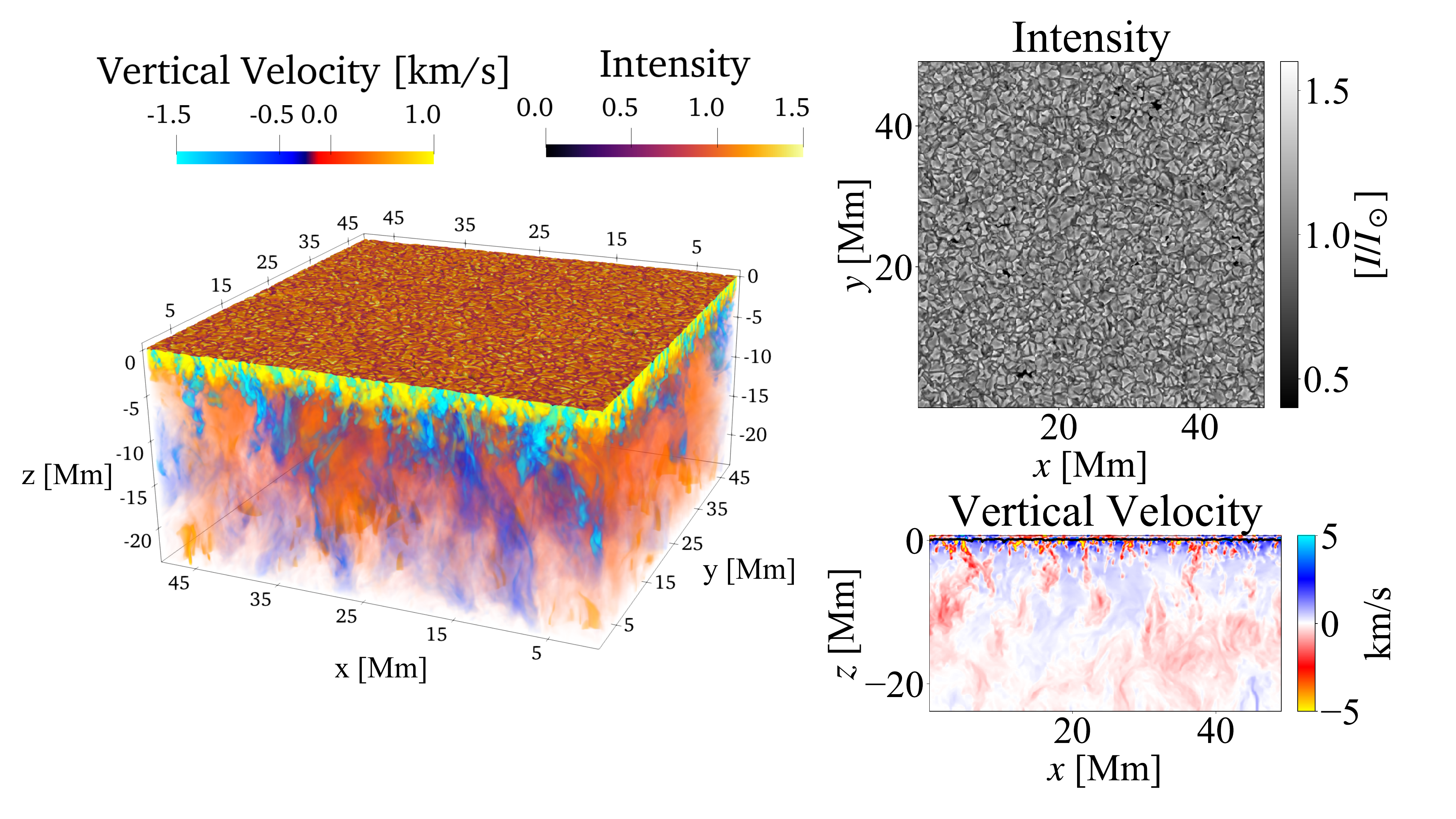}
  \end{center}
 \caption{Snapshots of a simulation result are shown. The left panel shows a three-dimensional volume rendering of the simulation. The color in the box represents the vertical velocity. The surface map on the box shows the radiative intensity. The panels on the right panels show the emergent radiative intensity (top) and a horizontal slice of the vertical velocity (bottom).
  {Alt text: A mesh-like granulation pattern is on the surface. In deeper areas in the Sun, the convective velocity decreases, and the spatial scale of the structures becomes larger.}
 \label{fig:simulation}}
 \end{figure*}

For the generality of the training data, the simulations are performed in five different initial magnetic fields. The snapshots of the simulation are shown in Figure \ref{fig:simulation}.

\subsection{Data processing}
In this study, we plan to apply our neural network to observational data taken from SDO/HMI. To fill the gap between the observation and the simulation, we carry out several data processing steps for the obtained data. The resolution of SDO/HMI is 0.5 arcsec, corresponding to 350 km at the solar disk center. The simulation data reproduce the observation data by applying SDO's PSF(point spread function) from \cite{Yeo+2014}. 2014 to the intensity, velocity field, and magnetic field. And the grid spacing of our simulation is 96 km, and we average 4$\times$4 grids into a grid to decrease the resolution mismatch. This procedure results in an image with a grid spacing of 384 km and a horizontal number of grid points of 128$\times$128.
\par
%In addition to the grid averaging, a Gaussian filter is applied to the obtained simulation data so that the power spectra of the observational and simulation data match each other. The widths of the Gaussian filter are $\sigma = $420, 417, and 192 km for the radiative intensity, the line of sight (LoS) velocity field, and the LoS magnetic field, respectively. These values are determined to minimize the squared error between the power spectra of the observational and the simulation data. A Gaussian filter is applied to the 128$\times$128 image.
%s, followed by a Fourier transform, and the power spectrum is created by binning the wavenumbers into 100 bins for comparison.}
\par
In addition, similar to \cite{masaki_2023PASJ...75.1168M}, the power spectrum of simulation and observation data indicats the small-scale structures in the observation data do not match the simulation data due to the observational noise. To reduce the negative effect of this power spectrum mismatch on the learning process, we add random noise to 5\% of the standard deviation of the Fourier-transformed quantities. This noise eliminates small-scale structures with high wavenumbers. Then, the inverse Fourier transform is applied to the data, which is used as the training data.

\subsection{Training for neural network}
We use 128$\times$128 pixel images (radiative intensity, LoS velocity, and LoS magnetic field at $\tau=1$ surface) for training as input data. Each of these inputs consists of three images taken at one-hour intervals. Thus, nine images of physical quantities with a resolution of 128$\times$128 pixels are input data. The LoS velocities at 5.4, 12, and 18 Mm depths from the surface with 512$\times$512 pixel images without any data processing are used as output data. Out of the five simulations performed with different initial conditions for the vertical magnetic field, two are used for actual training, one as validation data to monitor the training process, and the other as test data to evaluate the final network performance. Using the results of a different numerical code from the simulation code that is not used for the test data is more appropriate. However, another code is currently unavailable in our group. Therefore, the evaluated values in this study are considered as reference values. We use the intensity, vertical velocity, and vertical magnetic fields obtained from the simulation as input data.
\par
\begin{figure*}
  \begin{center}
   \includegraphics[width=\textwidth]{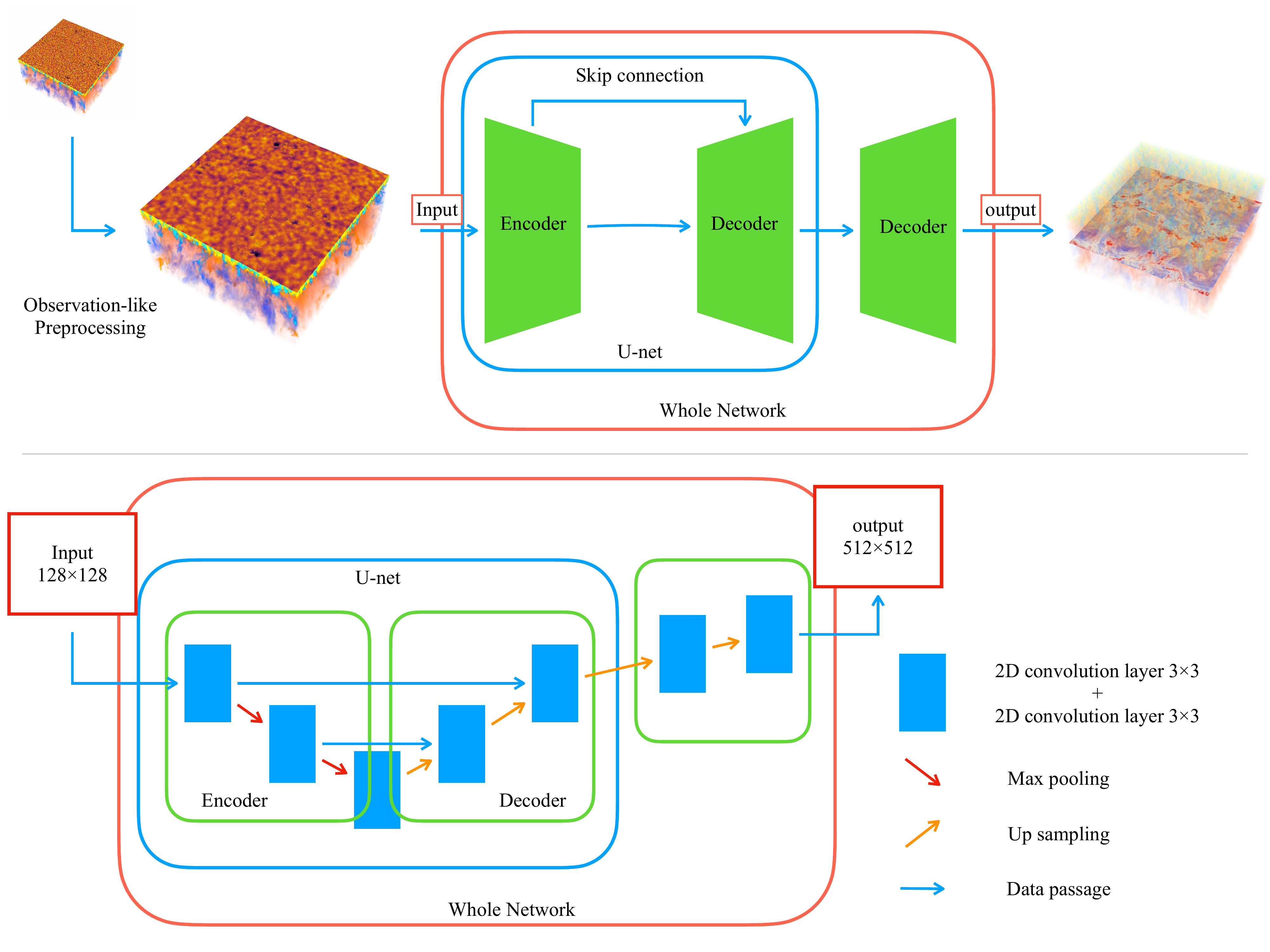}
  \end{center}
 \caption{ This figure shows the network architecture. The top diagram is the network's overall structure based on the U-net model used in this study. The bottom presents a detail of our structure. The blue squares represent the convolution layers. The U-net structure processes input data, two max pooling (shown by red arrows), and two upsamplings (shown by orange arrows). Then, it is upsampled again to match the resolution of the simulation.
  {Alt text: We plan to release the source code, including the network structure. Please refer to it for more details.}
 }\label{fig:training}
 \end{figure*}

The network architecture is shown in Figure \ref{fig:training}. This network structure is based on U-net \citep{ronneberger_2015arXiv150504597R}, which has an encoder-decoder architecture that efficiently processes images. However, while typical encoder-decoder structures usually have the same input and output image sizes \citep{masaki_2023PASJ...75.1168M}, in this study, the pixel numbers of the input data (128$\times$128) and the output data (512$\times$512) are different. The output data is made to the simulation as closely as possible, so we use the raw simulation data. Thus, the decoder structure is extended to match the size of the output images. As a result, the output has a higher resolution than the input, and the network can be considered super-resolution.
In the encoder, each convolutional layer allies two convolutions with a kernel size of 3×3. This is followed by downsampling using max pooling in a 2×2 region. The number of kernels doubles with each downsampling, starting with 256 layers in the first layer. In the decoder, after similar convolutions, upsampling is performed by expanding one pixel into a 2×2 region. Skip connections are applied by concatenating tensors along the kernel dimension.
Due to the our GPU performance limitations, the U-net is shallower than that of \cite{ronneberger_2015arXiv150504597R}.
The mini-batch size is four. Training is performed for eight epochs, and the model with the highest correlation coefficient on the validation data is chosen. We use mean squared error (MSE) as the loss function.
We use Adam optimizer \citep{kingma_2014arXiv1412.6980K} and the hyperparameters is same in the \cite{kingma_2014arXiv1412.6980K}. We use Approximately 120,000 data sets for the training data, while 1,000 data are used for validation.
This study uses TensorFlow, and the GPU is an Nvidia RTX 2080 Ti.

\section{Result}

\subsection{Comparison with simulation}
\begin{figure*}
  \begin{center}
   \includegraphics[width=\textwidth]{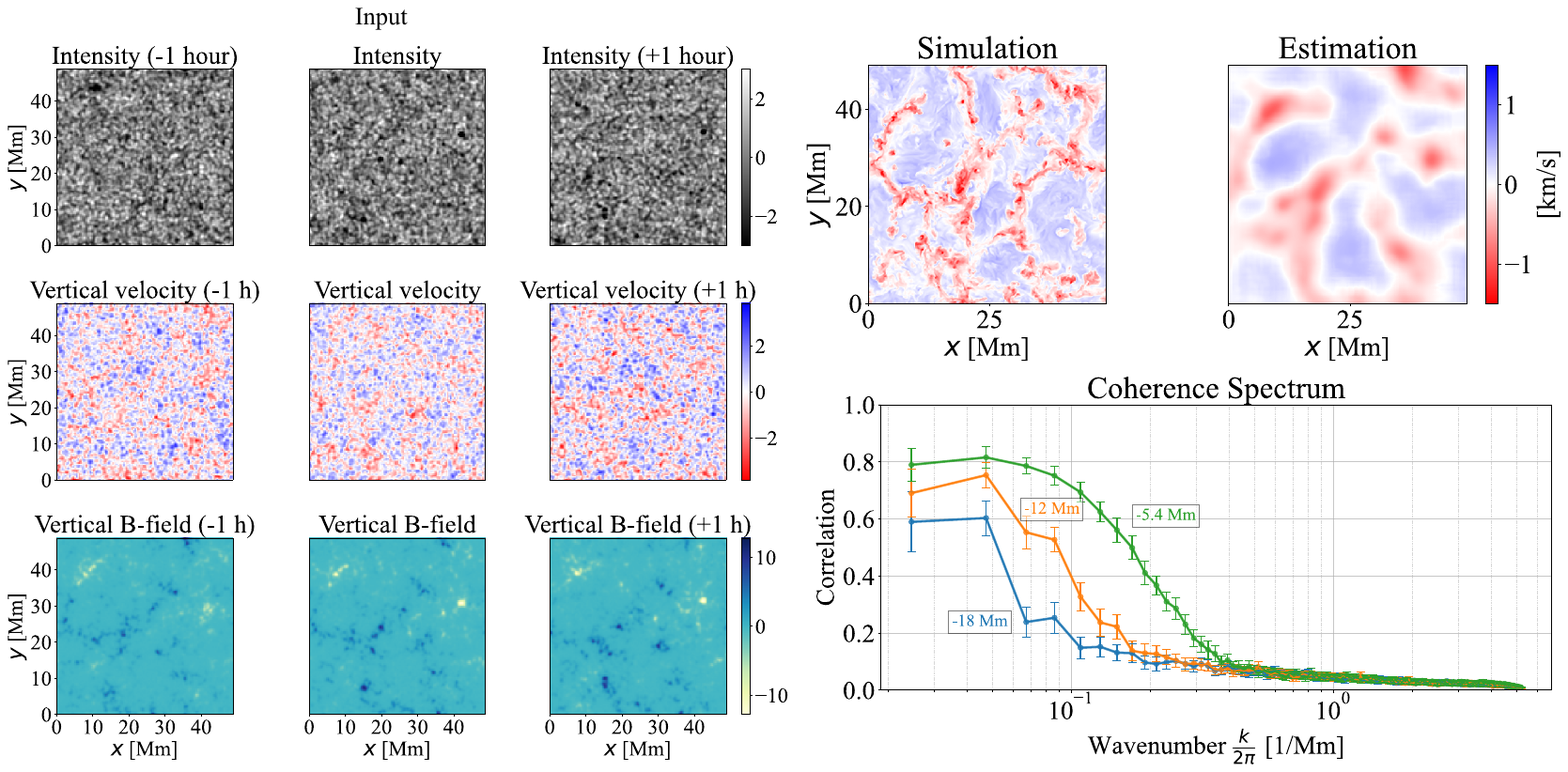}
  \end{center}
 \caption{The input data and a comparison between the estimation and simulation are shown. The left panel shows the input data for the neural network. These images represent the values normalized and processed. The top right panels show the result of the simulated and the estimated velocity fields by our neural network. The bottom right panel shows the coherence spectrum between the estimated and simulated velocities in the deep convection zone. The error bars represent the 99\% confidence interval.
  {Alt text: The coherence spectrum indicates that the values generally decrease in deeper regions, with the lines remaining aligned without crossing. It forms an S-shaped curve centered around 5 Mm, with values nearing zero at depths below 3 Mm.} }\label{fig:estimation}
 \end{figure*}

The left nine panels of Figure \ref{fig:estimation} show the input data for the network. The top, middle, and bottom rows show the radiative intensity, the LoS velocity field, and the LoS magnetic field at $\tau=1$ surface, respectively. The middle column shows the data at the same time as the estimated velocity field in a deep layer, and the left and right columns show the data one hour earlier and later, respectively. Each dataset is normalized to have a mean of 0 and a standard deviation of 1 to improve training efficiency.
\par
The top right two panels of Figure \ref{fig:estimation} show the LoS velocity at 12 Mm depth from the surface. The right and left panels show the numerically simulated and the estimated data by our neural network. Although small-scale structures are smoothed out, the large-scale convective structures are well reproduced. The correlation coefficients between the simulated and the estimated values are 0.25 at a depth of 5 Mm, 0.37 at a depth of 12 Mm, and 0.37 at a depth of 18 Mm. The lower correlation coefficient at 5 Mm is due to the difficulty in matching small-scale structures because the scale of convective structures is smaller than that of 12 Mm. The reproduction of the large-scale structure can be evaluated with the coherence spectrum, which is an indicator of the agreement of image structures at each scale, with values close to 1 indicating high agreement and values close to 0 indicating no correlation. See \citep{Ishikawa+2022} for the detailed definition of the coherence spectrum. The bottom left panel of Figure \ref{fig:estimation} shows the coherence spectra at each depth.
% The coherence spectrum is obtained as follows:
% Perform a two-dimensional Fourier transform on each of the two images to obtain X and Y.

% \begin{equation}
% \hat X(k_x, k_y)
% =
% \frac{1}{N^2}
% X(x, y)
% \exp[-i(k_xx+k_yy)]
% \end{equation}
% Calculate the sum of $X^2$ and $Y^2$ over an interval with a width of $\Delta k$ around a given k to obtain the power spectrum in that interval.

% \begin{equation}
% E_X(k)=\frac{1}{2\Delta k}
% \sum_{\sqrt{k_x^2+k_y^2}\in [k-\Delta k/2, k+\Delta k/2]}
% \left|\hat X(k_x, k_y)\right|^2
% \end{equation}
% Similarly, calculate the sum of $XY$ over the same interval to derive the cross-spectrum.

% \begin{equation}
%   S_{XY}(k)=\frac{1}{2\Delta k}
% \sum_{\sqrt{k_x^2+k_y^2}\in [k-\Delta k/2, k+\Delta k/2]}
% \left<\hat X^* \hat Y\right>
% (k_x, k_y)
% \end{equation}
% Calculate the coherence spectrum using the following equation.

% \begin{equation}
%   \gamma_{XY}(k)=
% \frac
% {S_{XY}(k)}
% {\sqrt{\left<E_X\right>(k) \left<E_Y\right>(k)}}
% \end{equation}
We divide the wavenumber into 256 bins and choose $\Delta k$ as an interval of $135.36~\mathrm{km^{-1}}$. From Figure \ref{fig:estimation}, we can see that those large-scale structures match with a value of about 0.7.
This result also aligns with the intuition that estimation becomes more difficult at greater depths.

\subsection{Comparison with Helioseismology}
We compare the detection performance of the internal structure estimated by our neural network with helioseismology, a method for probing the internal structure of the Sun. We adopt a helioseismology method suggested by \citet{GizonBirch2004}. Performing an inversion to calculate internal velocities adds new uncertainties, and we currently cannot carry out such types of calculations. In this study, we only calculate travel times with phase speed filters. We evaluate the points to annular travel time and can detect the divergent and convergent flows, which physically correspond to up- and downflows in the deep convection zone. Thus, we can compare these with our estimated velocity field. The distance between two points is 12 Mm, and we use the filter corresponding to 8.7-14.5 Mm, as shown in Table 1 on page 30 of \citet{GizonBirch2004}. The travel time shifts are calculated in 256 directions, evenly spaced over 360 degrees.
\par
\begin{figure*}
  \begin{center}
   \includegraphics[width=\textwidth]{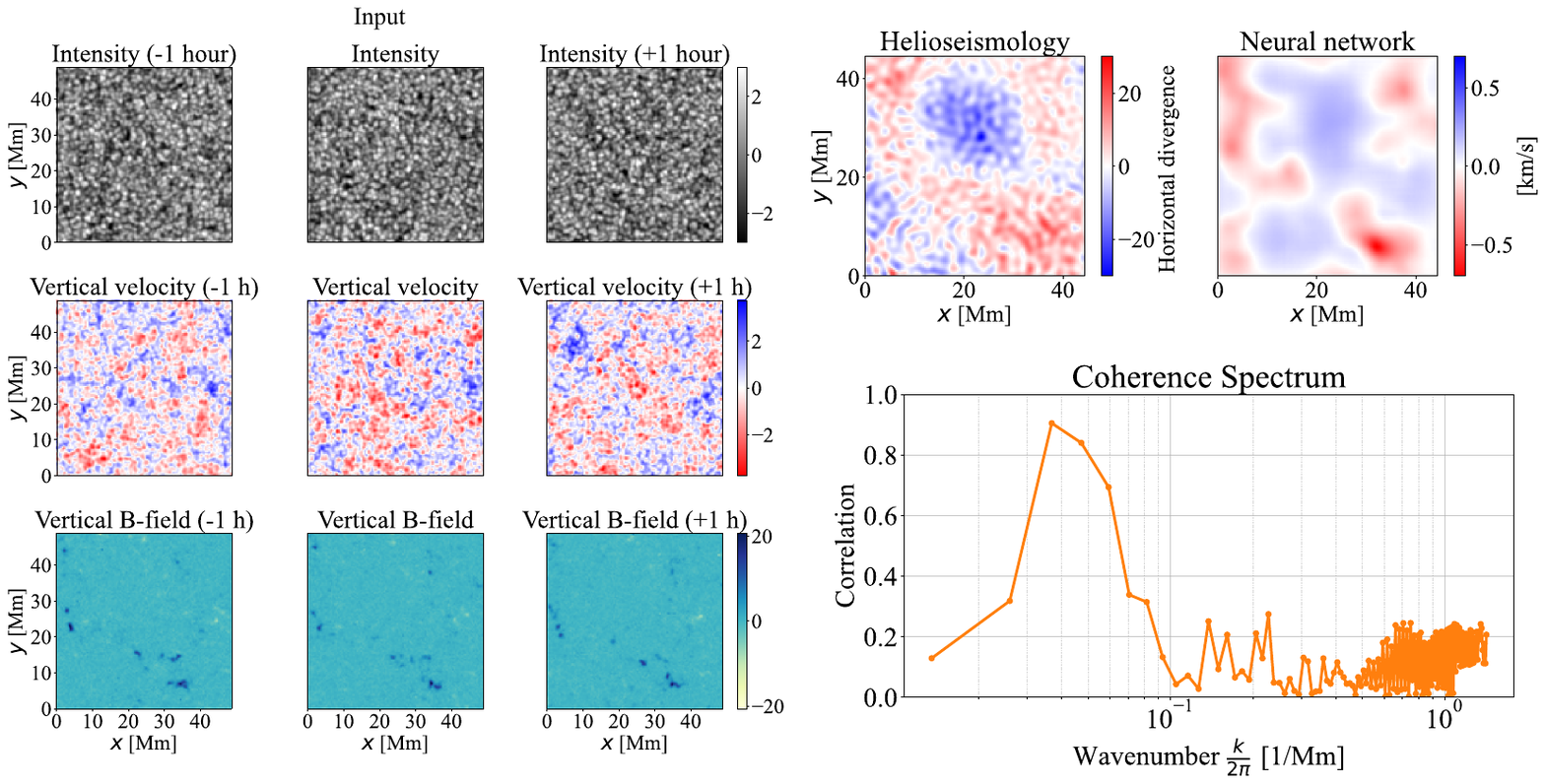}
  \end{center}
 \caption{Helioseismology result and network estimation from observation data. The left panels show the normalized values of the observed data input into the neural network, as shown in Figure \ref{fig:estimation}. The top right panel compares the results from the helioseismology and the estimated velocity field. The bottom right panel depicts the coherence spectra between helioseismology and the estimation.
 {Alt text: The estimation results from the network and the helioseismology results are in good agreement with the overall structure. In helioseismology, regions with lower travel time shifts are scattered on spatial scales of about 1 Mm, but such regions do not appear in the neural network results. The coherence spectrum has a shape similar to the simulation vs estimation graph in Figure \ref{fig:estimation}, but it is more irregular due to the smaller amount of data used.}
 }\label{fig:Helio_obs}
 \end{figure*}

In this study, we show a comparison between the helioseismology method and the proposed neural network for observational data. Although the same process was also conducted for the simulation data, it was not explicitly shown. The result for simulated data is similar to the following comparison for the observational data. Figure \ref{fig:Helio_obs} shows the results of comparison with the helioseismology for observational data. The data are from SDO/HMI on April 5, 2011, from 00:00:45 to 23:59:45. The sample of this data is shown in the same format as Figure \ref{fig:estimation} on the left side of Figure \ref{fig:Helio_obs}. The results use 1919 images from a day's worth of 45-s cadence data. Travel time is calculated with a window interval corresponding to 12 Mm. Note that this is a comparison made using one case's data, and it is not an error in the validated values that is not at a sufficient level of accuracy.
%, and the horizontal divergence is computed. Convergence corresponds to downflows, and divergent flows correspond to excessive upflows.
\par
The top right panels of Figure \ref{fig:Helio_obs} show a comparison between the travel time shift by the horizontal divergence (left) and the vertical velocity evaluated with our neural network (right). Because helioseismology estimates temporally average values, the network estimates are also averaged over time.
%The average is taken from 22 sets of estimated data from data of a day (24 images) at one-hour intervals (because three-hour sets are needed for estimation, the edges cannot be estimated at the same time).
Large-scale structures seem to match well. The correlation coefficient is -0.48, with a negative correlation due to the relationship between the travel time and the velocity field signs, i.e., the negative travel time shift corresponds to the horizontal divergent flow and upflow ($v_z>0$). The bottom right of Figure \ref{fig:Helio_obs} shows the coherence spectra for both methods. At scales around 20 Mm, the coherence is approximately 0.8, indicating a nice matching between the two.
%However, there is no match on smaller scales

\section{Summary and Discussion}
In this study, we propose a novel method to evaluate the solar internal flows. We train a neural network to estimate the velocity of upflows in the solar interior from three observable surface quantities: radiative intensity, LoS velocity, and LoS magnetic field. The training data are generated using the radiation magnetohydrodynamics simulation code R2D2, which reproduces internal convection in the Sun. Even with the small amount of data for the input, we nicely reproduce the overall velocity structure in the deep layer.
\par
A comparison with helioseismology using observational data shows a good correlation on a large scale, indicating that the estimation of the network velocity field does not significantly contradict reality. The neural network reproduces small-scale structures slightly better than helioseismology using the simulation data. This method could be better with small-scale structures. Even if the resolution of the output images is reduced, similar predictions can be made for large scales. This may be beneficial in terms of computational speed and data efficiency. However, even with high-resolution data for training, the accuracy does not improve. Drastic performance improvements are not expected with the current method, even with advances in observational technology.
\par
%The estimation becomes easier in shallower regions due to the stronger association with surface physical quantities, but it becomes difficult to estimate effectively at smaller convective scales.
In this study, a one-hour time interval is used for the input. The typical convection time and propagating time to the observable surface should depend on depth. Thus, there may be an optimal time interval for the target depth. Although using a higher cadence for data input can improve performance, increasing the data about 10 times will not lead to drastic improvements whereas it increases computational costs. Thus, appropriate input must be searched according to the purpose. This study uses simulations with resolution matched to SDO/HMI data as input data. Separately, even when training using the raw simulation data as input, the correlation coefficient does not significantly improve at depths shallower than 5 Mm. Therefore, even if observational technology advances and the accuracy of observational data improves, drastic performance improvements cannot be expected with this method.
\par
We apply our neural network for disk center data in this study. When we extend our method to other regions, solar global internal flow can be estimated, and we can possibly predict the emergence of rising flux tubes in the future.

\begin{ack}

\end{ack}

\section*{Funding}
This work was supported by JST SPRING, Grant Number JPMJSP2109.
The results were obtained using Cray XC50 at the Center for Computational Astrophysics, National Astronomical Observatory of Japan. This work was supported by MEXT/JSPS KAKENHI (grant no. JP23K25906, JP21H04497, and 21H04492) and by MEXT as "Program for Promoting Researches on the Supercomputer Fugaku" (Elucidation of the Sun-Earth environment using simulations and AI; Grant Number JPMXP1020230504).

\section*{Data availability}
The simulation code and data underlying this article are shared on reasonable request to the corresponding author. The machine learning code and trained models are publicly available at https://github.com/HiroyukiMasaki/Solar\_Interior.
% Sample Data Availability Statements
% https://academic.oup.com/pages/open-research/research-data#Data%20Availability%20Statements

% Any journal's BST file (e.g., apj.bst) can be used as PASJ's BST is unavailable.
% \bibliographystyle{****}
\bibliographystyle{plainnat}
\bibliography{reference.bib}

\end{document}